\documentclass[preprint,12pt]{elsarticle}

\usepackage{makecell} 

\usepackage{amssymb}
\usepackage{amsmath}

\usepackage{xcolor}

\journal{Acta Astronautica}

\begin{document}

\begin{frontmatter}



\title{Optimized Bragg Grating Liquid Crystal Solar Sail} 


\author[]{Sai Keshav Sasanapuri} 
\author[]{Xiulian (Nicole) Shan}
\author[]{Grover A. Swartzlander, Jr.\corref{cor1}}\ead{gaspci@rit.edu}
\cortext[cor1]{Corresponding author}

\affiliation[]{organization={Chester F. Carlson Center for Imaging Science, Rochester Institute of Technology},
            addressline={}, 
            city={Rochester},
            postcode={14623}, 
            state={NY},
            country={United States}}

\begin{abstract}
We have designed a thin 3-layer sun-facing diffractive Bragg reflector comprised of a chiral polymeric liquid crystal film. Using Rigorous Coupled-Wave Analysis (RCWA), the stack was optimized to maximize the transverse radiation pressure force across the wavelength band 0.47 - 1.67 $\mu \text{m}$.
The optimized structure operates on a single polarization state of light. A two structure stack, each operating on a different state of polarization, is predicted to provide
a transverse momentum transfer efficiency of 51\%.
This study demonstrates how a wavelength selective Bragg grating may be used both as a sunshade to scatter sunlight away from the Earth, and as a solar sail for maintaining a prescribed orbit (such as the Earth-Sun L1 point) owing to the momentum transfer efficiency. 
\end{abstract}



\end{frontmatter}



\section{Introduction}
\label{introduction}

Solar radiation pressure has been a subject of interest since the early work of Kepler in 1619, who proposed it as an explanation for comet tails \cite{Kepler1619}. With the development of electromagnetic theory, it was established that light carries momentum and can exert pressure on matter, forming the physical basis of solar sailing.
Unlike conventional propulsion systems, solar sails require no onboard propellant, relying instead on the 
continuous momentum transfer of scattered light
which affords continuous acceleration and high 
$\Delta V$ missions 
\cite{McInnes1999,Vulpetti2014, friedman2015human, johnson2023solar}. Harnessing solar radiation pressure for in-space propulsion remains a highly active area of research \cite{johnson2025solar}.
Conventional mirrored reflective solar sails generate radiation pressure primarily normal to the surface, thereby requiring tilting away from the sun to achieve a significant component of force transverse to the sun line.
When simultaneously used as a sunshade, a tilted mirrored sail decreases the amount of insolation.  What is more, navigation by tilting a large area sail can impose structural and control difficulties.
Diffractive solar sails have emerged as a promising alternative in which propulsion is achieved through engineered diffraction rather than specular reflection \cite{Swartzlander2017_diffractive_sail, swartzlander2018flying, Swartzlander2022}

\begin{figure}[h]
\centering\includegraphics[width=0.65\linewidth]{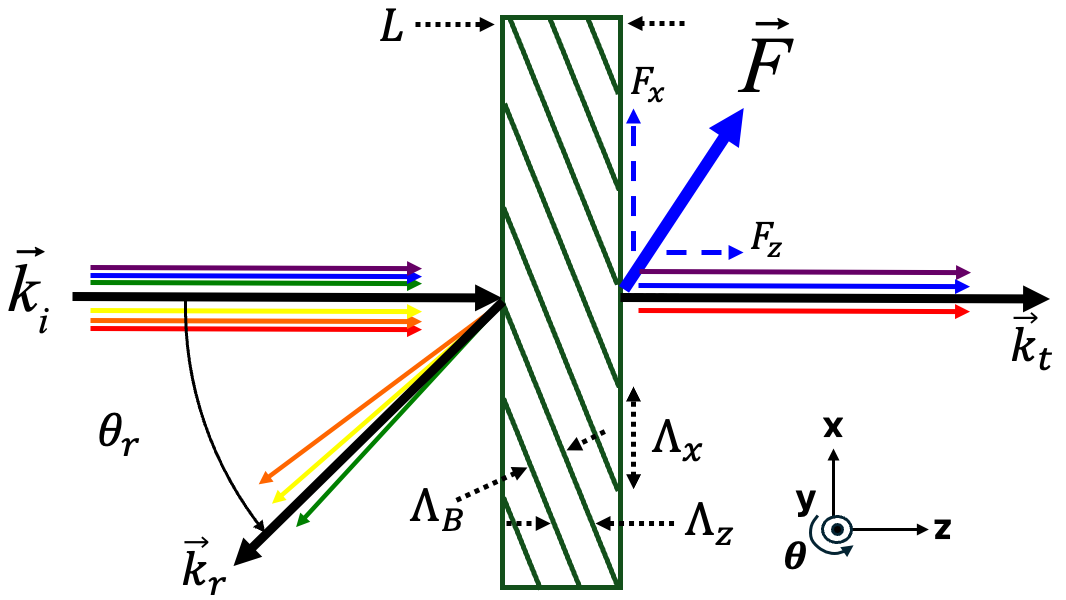} 
\caption{A sun-facing polarization Bragg grating solar sail having incident wave vectors $\vec{k}_i$ (for one polarization state) reflects wavelengths near the Bragg condition $(\vec{k}_{r})$ and transmits others $(\vec{k}_{t})$, resulting in a net radiation pressure force $\vec{F}$.
The transverse component $F_x$ is primarily important for navigation in a solar orbit.
The periodic structure is characterized by the Bragg period $\Lambda_B$, and resulting transverse and longitudinal periods $\Lambda_x$ and $\Lambda_z$, respectively.  The structure has a thickness $L > \Lambda_z$.  The reflection scattering angle $\theta_r$ satisfies the grating equation
$| \sin\theta_r | = \lambda / \Lambda_x$.
}
\label{fig:diffractive_kinematics}
\end{figure}

As illustrated in Fig.\ref{fig:diffractive_kinematics}, by 
scattering light via Bragg reflection,
these thin films can generate significant transverse radiation pressure force $F_x$ while maintaining a sun-facing orientation. 
Bragg gratings
(also called tilted volume hologram)
achieve extraordinarily large diffraction efficiencies over a modest wavelength bandwidth owing to simultaneously
satisfying both the grating equation,
the law of reflection from tilted periodically modulated refractive index structures, and constructive interference of reflected waves.  The Pancharatnam phase \cite{pancharatnam1956phenomenological} in anisotropic materials such as cholesteric liquid crystal polymers affords large refractive index modulations that self-align \cite{john1995bragg, sutherland2007liquid}.
Other methods of scattering light from a sun-facing film include metallic-coated diffraction gratings and hybrid reflection/transmission designs \cite{Crum2023_metallic,Crum2024_hybrid}. 
However, liquid-crystal-based diffractive optics are particularly promising owing to 
their thin, robust yet flexible properties
as well as scalable manufacturability
\cite{Roberts2023,Tabiryan2025}. 

Polarization gratings and polarization volume gratings provide engineering latitude for controlling light \cite{Gori1999,Lajunen2004,Oh2008,Tabiryan2016,Roberts2019}. 
In particular, polarization volume gratings (PVGs) can achieve near-unity diffraction efficiency and large diffraction angles under Bragg conditions \cite{Weng2016_PVG}, while advanced diffractive waveplate designs allow polarization-independent operation and enhanced control of diffraction behavior \cite{Roberts2018_DW}. Bragg polarization gratings have also been demonstrated to provide high efficiency over wide angular bandwidths, making them attractive for high-performance optical redirection systems \cite{Xiang2018_BraggPG}.

Accurate modeling of these complex periodic and anisotropic structures requires rigorous electromagnetic methods. Rigorous coupled-wave analysis (RCWA) provides an efficient and exact framework for solving diffraction from periodic gratings, including anisotropic and multilayer configurations \cite{Moharam1981}. Extensions of RCWA to liquid crystal polarization gratings enable detailed analysis of diffraction efficiency, polarization response, and transitions between Raman–Nath and Bragg regimes \cite{Xiong2020_RCWA_LCPG}. Combined with transfer-matrix approaches for birefringent media \cite{Yeh1979,Rokushima1983}, these methods provide essential tools for designing and optimizing high-efficiency diffractive optical systems.

In this work, we investigate a multilayer diffractive solar sail based on Bragg reflection using chiral liquid crystal structures. The proposed design employs stacked birefringent layers to achieve broadband spectral coverage while maximizing transverse momentum transfer efficiency. By combining polarization grating concepts with Bragg-type multilayer architectures, the structure enables efficient light redirection without requiring mechanical tilting. We determine diffraction efficiencies using the RCWA method for an anisotropic material, and then calculate the radiation pressure force expected from a solar blackbody. Our results demonstrate that optimized multilayer liquid crystal diffractive structures can achieve high transverse radiation pressure while maintaining a Sun-facing orientation, offering a promising pathway toward advanced solar sail propulsion systems that can also function as wavelength selective sunshades.

\section{Theoretical Framework and Methodology}
\label{theory}

\subsection{The Bragg Condition}
\label{bragg condition}
\begin{figure}[h]
\centering\includegraphics[width=0.7\textwidth]{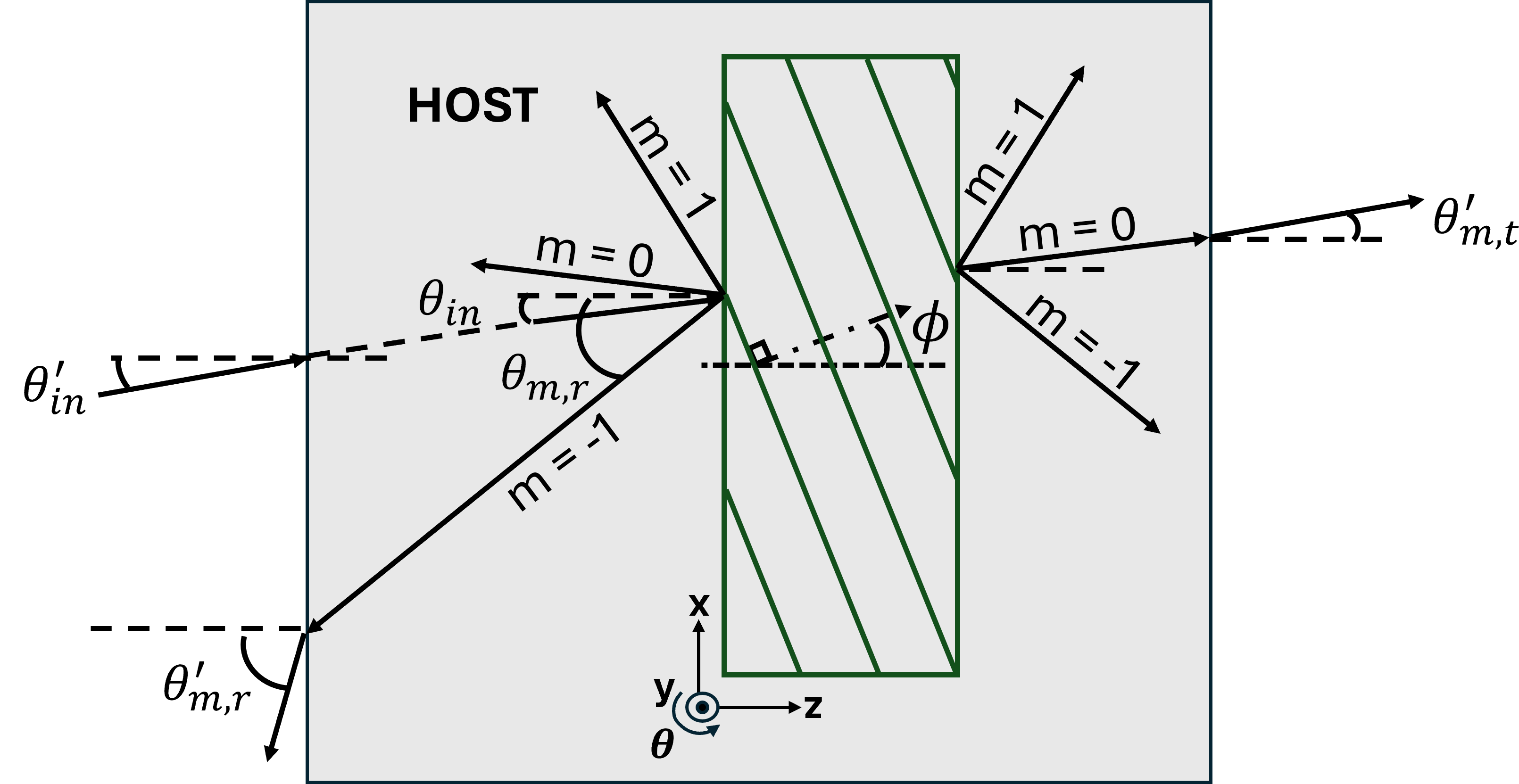}
\caption{Modulated refractive index structure, tilted at the angle $\phi$, embedded into an
index-matched host material (gray).  Incident sunlight at angle $\theta_\text{in}'$ is refracted into the
host at $\theta_\text{in}$ and scattered into reflected and transmitted diffraction orders.  Those orders
refract back into the vacuum.
}
\label{fig:grating_geometry}
\end{figure}

The fundamental structure of a Bragg grating is illustrated in Fig. \ref{fig:grating_geometry}
which shows titled lines of modulated refractive index at the angle $\phi$ with respect to the 
incident surface normal.  For convenience we depict the grating embedded in an index matched
homogeneous medium having refractive index $n_\text{eff}$.  According to Snell's law, 
a ray refracts from the angle $\theta_\text{in}'$ in vacuum to the angle $\theta_\text{in}$ in the matched medium
(i.e., $\sin\theta_\text{in}' = n_\text{eff} \sin\theta_\text{in}$).  Owing to the periodicity of the grating one may expect
multiple diffraction orders $m$ to be reflected and transmitted.  The $m=0$ modes correspond
to reflection and transmission of an unmodulated structure.  The $m=-1$ reflection order corresponds to
the law of reflection from the tilted modulation lines.  When such reflections constructively add, the net
reflected electric field approaches that of the incident field.  That net electric field refracts back
into the vacuum at the angle $\theta_{-1,\text{r}}$ \textcolor{blue}{$\theta_{-1,\text{r}}^{'}$}.  As described below the transverse momentum 
transfer efficiency approaches 100\% when both the reflection into the vacuum approaches 100\% 
and $|\theta_{-1,\text{r}}|$ approaches $90^\circ$.  To optimize these two parameters for sunlight
the grating must account for the momentum (or energy) irradiance distribution of the incident light.
Below we assume a blackbody distribution.  

The Bragg condition relates the modulation period $\Lambda_B$ with the slant angle $\phi$, 
design wavelength $\lambda_b$, incidence angle $\theta$, and, refractive index $n_\text{eff}$:
\begin{equation}
\Lambda_{B} = \frac{\lambda_{b}}{2 \cdot n_{eff} \cdot \cos(\phi + \theta_{in})}
\label{eq:bragg_condition}
\end{equation}
\noindent
For a uniaxial liquid crystal the effective refractive index is given by 
\begin{equation}
n_{eff} = \sqrt{ (n_e^2 + 2n_o^2)/3 \;} 
\end{equation}
\noindent
where $n_o$ and $n_e$ are respectively the ordinary and extraordinary refractive indexes of the liquid crystal \cite{Roberts2018_DW}. 
The grating wave vector of magnitude $K_B = 2 \pi/\Lambda_B$ may be expressed as components
of magnitude $2\pi/\Lambda_x$ and $2\pi/\Lambda_z$,
resulting in respective transverse and longitudinal periods 
$\Lambda_x = \Lambda_B / \sin\phi$ and $\Lambda_z = \Lambda_B / \cos\phi$.

\subsection{Solar Blackbody Irradiance Frequency Distribution}
\label{solar blackbody spectral irradiance}
The transverse radiation pressure force exerted on a solar sail may be optimized by accounting for the
peak of the solar frequency distribution, which may be estimated by the blackbody irradiance $I_{\nu}$ 
at the temperature $T = 5778$ K, which according to Planck's law may be expressed at a distance $r$ from the Sun:
\begin{equation}
    I_{\nu} = \left(\frac{R_S}{r}\right)^2 \; \frac{2\pi h \nu^3}{c^2} \; \frac{1}{e^{\frac{h \nu}{k_B T}} - 1}
    \label{eq:planck_law}
\end{equation}
\noindent where $\nu = c / \lambda$ is the optical frequency, $c$ is the speed of light, $h$ is the Planck constant,
$k_B$ is the Boltzmann constant,  and $R_S$ is the radius of the Sun.
The peak of this distribution occurs at $\nu_\text{peak} = 340$ THz.
 Note: the corresponding wavelength $\lambda_\text{peak} = 883$ nm differs from the peak of the
 wavelength-dependent distribution (502 nm) because $d\nu = -(c/\lambda^2) d\lambda$.
 An optimized Bragg solar sail should therefore be optimized in the vicinity of $\lambda_b = 883$ nm,
 or equivalently $\nu_b = c /  \lambda_b = 340$ THz.  As described below, multiple Bragg
 gratings may be stacked, with each optimized at a different region of the blackbody spectrum.
Such stacking can be achieved simply by changing the tilt angle $\phi$ and the design wavelength $\lambda_b$
for each layer, or by continuously chirping the structure.  To demonstrate underlying principles,
the former procedure is adopted in this report by use of three stacked Bragg gratings,
each comprised of the same material.
For numerical stability our optimization range covers 180 to 640 THz
(or equivalently 1.67 to 0.47 $\mu$m), encompassing 69.6\% of the total blackbody
energy flux of $I_\text{sun}=$ 1367 W/m$^2$ at 1 AU from the sun.
The momentum transfer efficiency of the sail may be benchmarked 
using either the irradiance per polarization, $(1/2) I_\text{sun}=$ 683.5 W/m$^2$ 
or the full solar constant, $I_\text{sun}$.

\subsection{RCWA Modeling and Momentum Transfer}
\label{rcwa modelling and momentum transfer}
A common numerical method of solving for optical diffraction from a periodic structure
is Rigorous Coupled-Wave Analysis (RCWA) \cite{Moharam1981}.
In our case the structure is comprised of a chiral liquid crystal polymer
which has advantages such as high index contrast (birefringence), 
flexible mechanical properties, and ease of large area fabrication.
Each polarization state diffracts independently of the other.
For example, a well designed grating diffracts right hand circularly polarized light (RHCP)
when the Bragg condition is satisfied, but the left circularly polarized state
remains unaffected.  Various methods may be used to diffract both
polarization states, but this is beyond the scope of this report.


For a given incident frequency of a given circular polarization state, CP,
the RCWA code outputs reflection and transmission values at
diffraction orders $m$, $R_{\nu,m}^{\text{(CP)}}$ and $T_{\nu,m}^{\text{(CP)}}$, respectively.
This allows the determination of the reflected and transmitted spectral irradiance values 
$I_{\nu,r,m}^{\text{(CP)}}  = \frac{1}{2} R_{\nu,m}^{\text{(CP)}} I_\nu$ and
$I_{\nu,t,m}^{\text{(CP)}}  = \frac{1}{2} T_{\nu,m}^{\text{(CP)}} I_\nu$, respectively.

\subsection{Momentum Transfer Efficiency}
The grating equation provides the diffracted scattering angles within the host material:
\begin{equation}
\sin\theta_m = \sin\theta_\text{in} 
    \textcolor{black}{-} m \lambda/n_\text{eff} \Lambda_x
			\label{eq:grating_equation_host}
\end{equation}
Applying Snell's law, the diffracted scattering angles in vacuum are given by
\begin{equation}
\sin\theta_m' = n_\text{eff} \sin\theta_m = \sin\theta_\text{in}' \textcolor{black}{-} m \lambda/ \Lambda_x
			\label{eq:grating_equation_vacuum}
\end{equation}
Care must be taken for the angle definitions when multiple gratings are stacked.
For example, the output angle from one layer becomes the incident angle for the next layer.

The momentum transfer efficiency distribution across frequency may be expressed
for the reflection and transmission modes
\begin{equation}\begin{split}
\vec{\eta}^{\, \text{(CP)}} (\nu) 
	& = \sum_m \frac{I_{\nu,r,m}^{\, \text{(CP)}} \big( \vec{k}'_\text{in} - \vec{k}'^{(r)}_m \big)}{I(\nu) k'} + 
	\sum_m \frac{I_{\nu,t,m}^{\, \text{(CP)}} \big( \vec{k}'_\text{in} - \vec{k}'^{(t)}_m \big)}{I(\nu) k'} \\
	& = \sum_m \frac{1}{2} \, R_{\nu,m}^{\, \text{(CP)}} \frac{ \vec{k}'_\text{in} - \vec{k}'^{(r)}_m }{ k'} + 
	\sum_m \frac{1}{2} \, T_{\nu,m}^{\, \text{(CP)}} \frac{ \vec{k}'_\text{in} - \vec{k}'^{(t)}_m }{ k'} 
			\label{eq:eta}
\end{split} \end{equation}
where $\vec{k}'_\text{in}$ is the incident wave vector measured in vacuum,
$\vec{k}'^{(r,t)}_m$ is the m$^\text{th}$ order scattered wave vector measured in vacuum,
and $k' = | \vec{k}'_\text{in} | = | \vec{k}'^{(r,t)}_m | = 2 \pi / \lambda$
is the magnitude of the wave vectors in vacuum,
where $\lambda = c / \nu$ is the vacuum wavelength.
The summations in Eq.\ref{eq:eta} apply only over the orders satisfying the cut-off condition
\begin{equation}
| \sin\theta_\text{in}' \textcolor{black}{-} m \lambda/ \Lambda_x | < 1
\end{equation}
The wave vectors may be decomposed into transverse and longitudinal components:
\begin{subequations} \begin{align}
\vec{k}'_\text{in} & = k' \big( \sin\theta'_\text{in} \; \hat{x} + \cos\theta'_\text{in} \; \hat{z} \big)
\\
\vec{k}'^{(r)}_m & = k' \big( \sin\theta'_m \; \hat{x} - \cos\theta'_m \; \hat{z} \big)
\\
\vec{k}'^{(t)}_m & =k' \big( \sin\theta'_m \; \hat{x} + \cos\theta'_m \; \hat{z} \big)
\end{align} \end{subequations}
where $\cos\theta_m' = \sqrt{1 - \sin^2\theta_m'}$ and $\cos\theta_\text{in}' = \sqrt{1 - \sin^2\theta_\text{in}'}$.
The transverse and longitudinal efficiency distributions may then be expressed (for a sail
operating on a single polarization of unpolarized light)
\begin{subequations} \begin{align}
\eta_{x}^\text{(CP)} (\nu) = &
\sum_m \frac{1}{2} R_{\nu,m}^\text{(CP)} ( \sin\theta'_\text{in} - \sin\theta'_m) +
\sum_m \frac{1}{2} T_{\nu,m}^\text{(CP)} ( \sin\theta'_\text{in} - \sin\theta'_m) \\
\eta_{z}^\text{(CP)} (\nu) = &
\sum_m \frac{1}{2} R_{\nu,m}^\text{(CP)} ( \cos\theta'_\text{in} + \cos\theta'_m) +
\sum_m \frac{1}{2} T_{\nu,m}^\text{(CP)} ( \cos\theta'_\text{in} - \cos\theta'_m) 
\end{align} \end{subequations}
Finally we integrate to obtain the net momentum transfer across a band of frequencies
from $\nu_\text{min}$ to $\nu_\text{max}$:
\begin{subequations} \begin{align}
\bar{\eta}_{x}^\text{(CP)} = &
	 \int_{\nu_\text{min}}^{\nu_\text{max}}  I(\nu) \; \eta_{x}^\text{(CP)} (\nu) \; d\nu  \Big/ I_\text{sun}
		\label{eq:etax}
\\
\bar{\eta}_{z}^\text{(CP)} = &
	 \int_{\nu_\text{min}}^{\nu_\text{max}}  I(\nu) \; \eta_{z}^\text{(CP)} (\nu) \; d\nu  \Big/  I_\text{sun}
		\label{eq:etaz}
\end{align} \end{subequations}
where $I_\text{sun} = \int_0^\infty  I(\nu) \; d\nu = 1367$ W/m$^2$.
In principle the frequency integration limits would be set to $0$ and $\infty$
but numerical integration requires bounds. Hence we set
$\nu_\text{min} = 180$ THz and $\nu_\text{max} = 640$ THz.

\section{Optimization Strategy}
To optimize Bragg diffraction for a given polarization state for the purpose of 
navigating in a solar orbit, e.g., a station keeping at the Earth-Sun L1 point,
we seek the parameters such as $\Lambda_B$, $\phi$, and $\lambda_b$ that
produce the greatest component of force perpendicular to the sun line 
$F_x = I_\text{sun} \bar{\eta}_x^\text{(RCP)} / c$.
Here we have arbitrarily selected the right circularly polarized state (RCP).

\subsection{Birefringence and Bandwidth}
\begin{figure}[h]
\centering\includegraphics[width=0.75\linewidth]{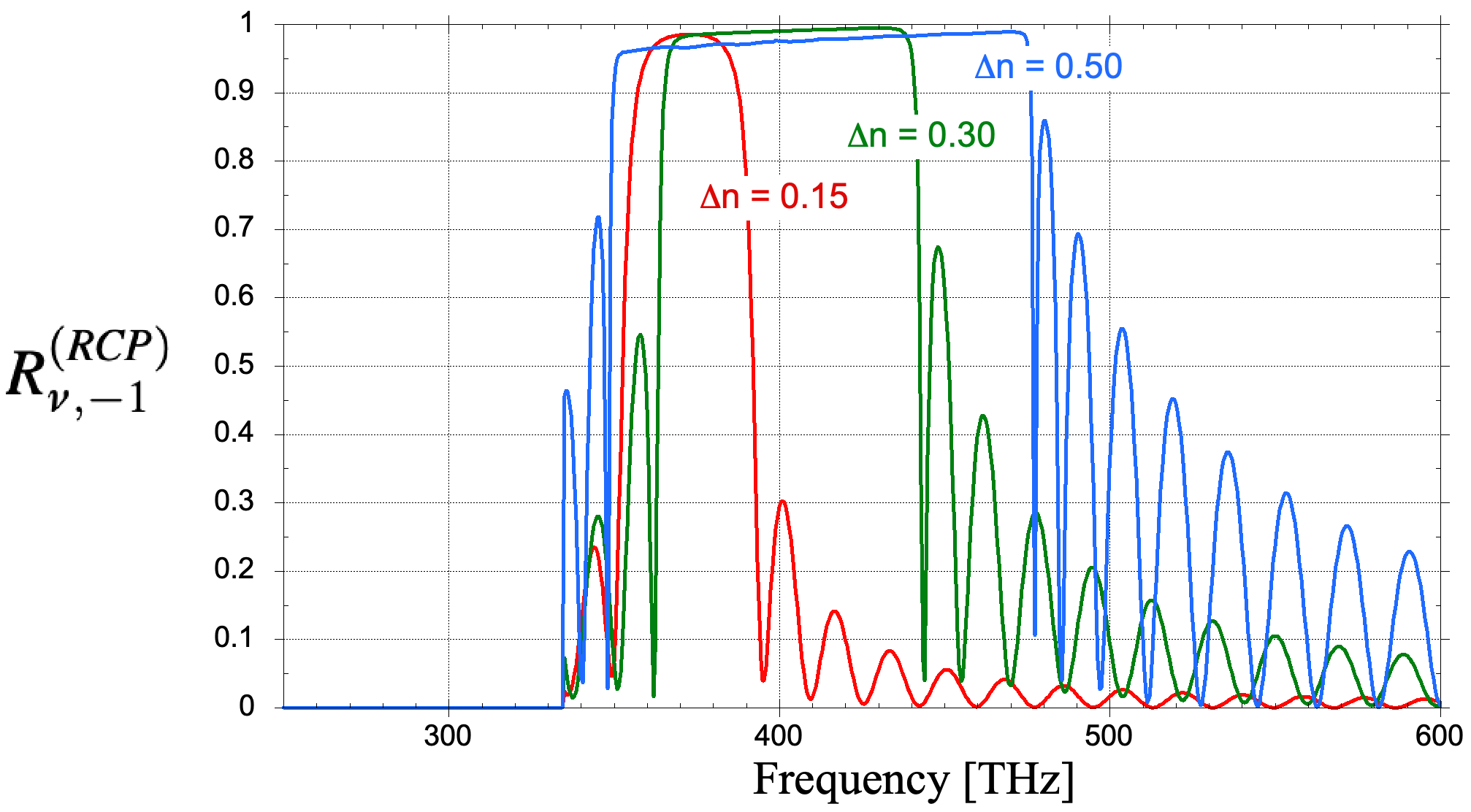}
\caption{Frequency dependent Bragg reflection coefficient $R_{\nu,-1}^\text{(RCP)}$ depicting increasing bandwidth
with increasing birefringence $\Delta n$.
}
\label{fig:birefringence_sweep}
\end{figure}

\begin{figure}[h]
\centering\includegraphics[width=0.6\linewidth]{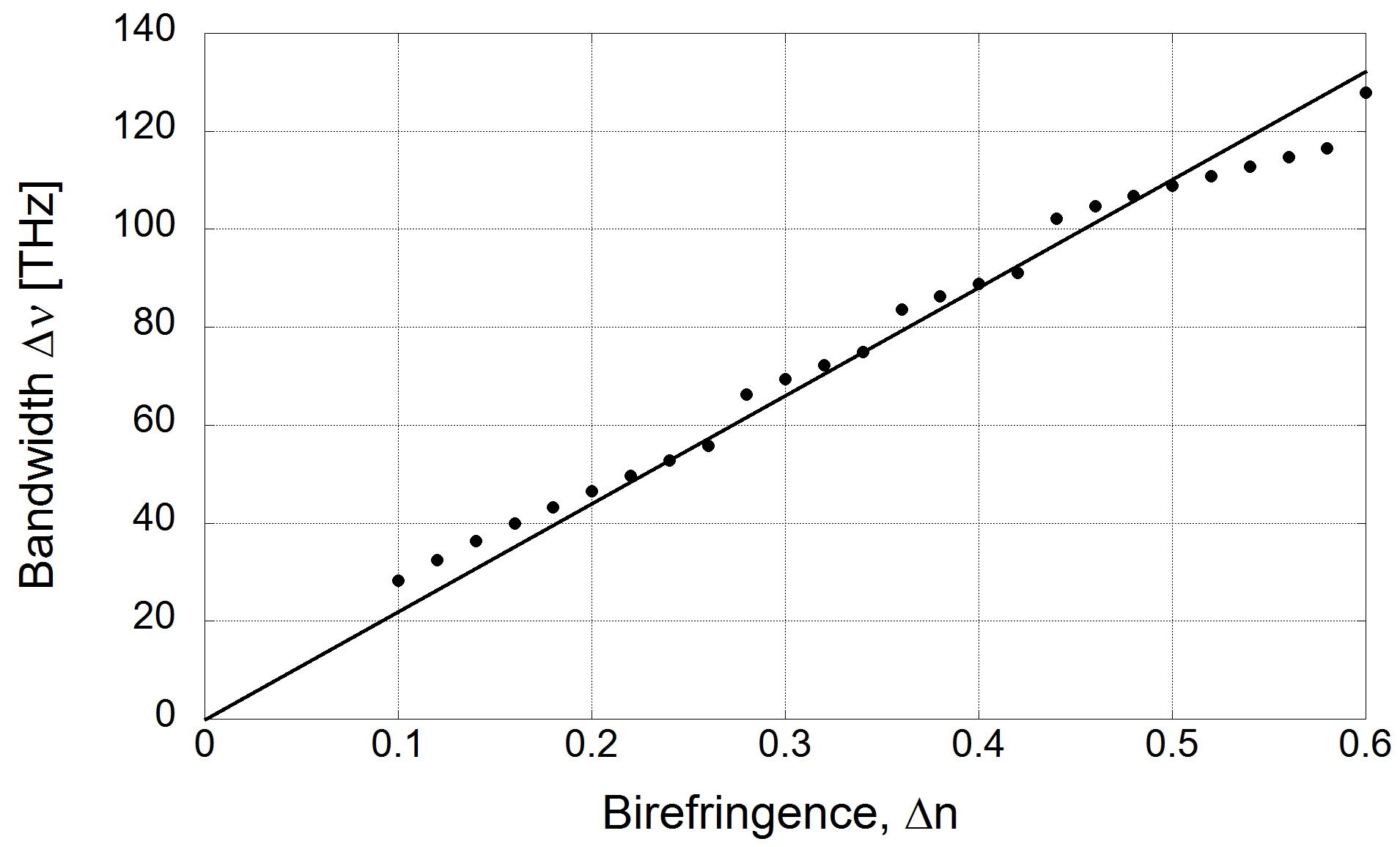}
\caption{
Full bandwidth $\Delta \nu$ at the $R_{\nu,-1}^{(RCP)} = 0.5$ value assuming normal incidence, slant angle $\phi=17.5^\circ$, and Bragg period $\Lambda_{\text{B}}=0.296\mu m$.
A straight line fit (solid black line) provides 
$\Delta \nu =(220\text{THz})\Delta n$.
}
\label{fig:FWHM_vs_Birefringence}
\end{figure}

The bandwidth of Bragg reflection is known to increase with increasing birefringence 
$ \Delta n  = | n_\text{e} - n_\text{o}|$, as illustrated in Fig.s 
\ref{fig:birefringence_sweep} and \ref{fig:FWHM_vs_Birefringence} for the case 
of normal incidence $\theta_\text{in}' = 0$, with
slant angle $\phi = 17.5^\circ$ and
Bragg period $\Lambda_B = 296$ nm
(the Bragg wavelength $\lambda_{\text{b}}$ and effective index $n_{\text{eff}}$ vary with birefringence).
In this case we numerically obtain a linear relation through
the origin: $\Delta \nu = ( 220 \text{THz} )  \Delta n $.
Below we assume a feasibly achievable birefringence \cite{gauza2005high} 
of $\Delta n = 0.5$ with $n_o = 1.7$ and $n_e = 1.2$ and $n_\text{eff} = 1.55$. 
These and other refractive index parameters are listed in Table \ref{tab:refractive_indices}.

\begin{table}[htbp]
\centering
\caption{Refractive indices used for the optical simulation and material modeling. \smallskip}
		\label{tab:refractive_indices}
\begin{tabular}{lc}
\hline
\textbf{Property} & \textbf{Value} \\
\hline
Ordinary Index ($n_o$) & $1.7$ \\
Extraordinary Index ($n_e$) & $1.2$ \\
Birefringence ($\Delta n$) & $0.5$ \\
Effective Host Index ($n_{eff} = n_{host}$) & $1.55$ \\
External Index ($n_{vacuum}$) & $1.0$ \\
\hline
\end{tabular}
\end{table}

\subsection{Three-Layer Bragg Grating Stack}
\begin{figure}[ht]
\centering\includegraphics[width=1\linewidth]{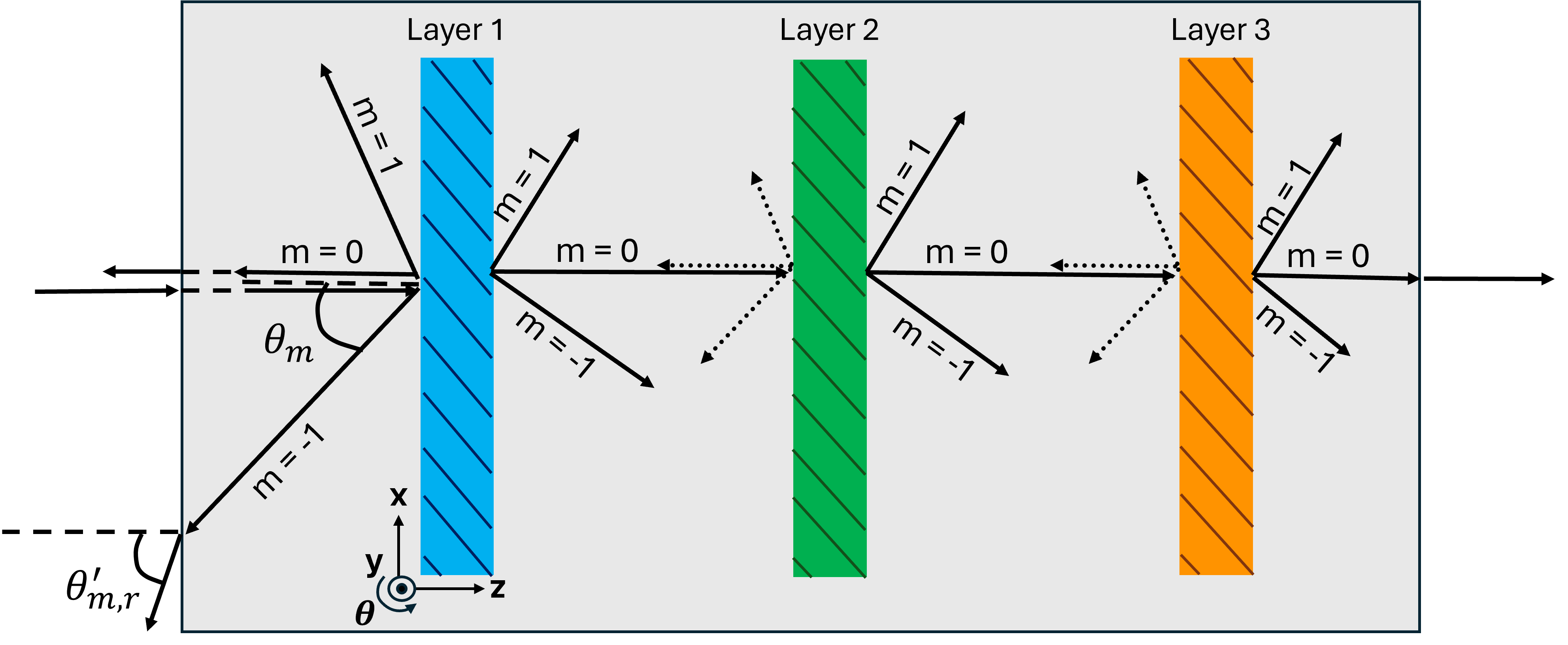}
\caption{Three-layer Bragg grating stack.  Layers 1 (blue) scatters short wavelength light.
Layer 2 (green) and Layer 3 (orange) scatter mid and long wavelength light, respectively.
Distances between layer is exaggerated to aid the eye.
}
\label{fig:multilayer_chaining}
\end{figure}

At $\Delta n = 0.5$ the grating bandwidth
of $\sim 115$ THz is small in comparison to the $\sim 500$ THz bandwidth of 
the solar spectrum, and thus, we propose stacking multiple Bragg
gratings, each optimized for a different region of the solar spectrum.
These bands may be placed adjacent to each other (as described below), 
or spectral gaps may be engineered to allow for the transmission of
unscattered light (as may be desired for sunshades that affect neither
photosynthesis nor photovoltaic power generation).
The layers are arranged such that the incident solar flux interacts with the 
highest-frequency grating first, as depicted in Fig. \ref{fig:multilayer_chaining}.

We assume right circularly polarized light is normally incident upon Layer 1 ($\theta_\text{in} = 0$), 
whereupon light is either primarily reflected into the $m=-1$ order, or transmitted 
into the $m=0$ order.  
The polarization is unchanged upon reflection and transmission.
Layer 2 receives the transmitted field from Layer 1, with the 
dominant incident mode also at normal incidence upon Layer 2.
Again light is either primarily reflected into the $m=-1$ order, 
or transmitted into the $m=0$ order.  Light reflected from Layer 2 does not
satisfy the Bragg condition of Layer 1, and thus it is assumed to pass through
Layer 1 unimpeded.  

A similar process occurs for Layer 3.
The dominant reflected waves therefore refract into the vacuum at angles satisfying the grating equation for each each layer:
\begin{equation}
\sin\theta_{-1}^{'(j)}  =  \lambda / \Lambda_x^{(j)}
\end{equation}
\noindent
For example, if $\sin\phi^{(j)} = 17^\circ$ for $j=1,2,3$, and 
$\Lambda_x^{(j)} = (0.69, 0.98, 1.45) \, \mu$m,
$\Lambda_B^{(j)} = (0.201, 0.296, 0.437) \, \mu$m,
and $\lambda_b^{(j)} = (0.60, 0.87, 1.29) \, \mu$m,
then at $\lambda = \lambda_b^{(j)}$ we obtain
$\theta_{-1}^{(j)} = $
$(60.5^\circ,$ 
$62.5^\circ,$ 
$62.9^\circ)$,
which are significantly large scattering angles.



\subsection{Optimization Algorithm}
The sequential optical response of the cascaded layers introduces severe non-linearities and 
inter-layer electromagnetic coupling. Because the spectrum transmitted by Layer 1 dictates the 
incident field for Layer 2, traditional gradient-descent algorithms are highly susceptible to becoming 
trapped in local minima. 
With the birefringence for each layer constrained, the search space for the three-layer stack spans 9 
continuous dimensions: the slant angles $\phi^{(j)}$, the cutoff frequencies $\nu_c^{(j)} = c/\Lambda_x^{(j)}$, 
and the thicknesses $L^{(j)}$. 
The layer thicknesses are individually tailored to minimize the mass of the Bragg sail. 
To robustly navigate the remaining parameter space, a Differential Evolution stochastic global optimization 
algorithm is employed. For a given parameter vector $[\phi_1, \nu_{c1}, \phi_2, \nu_{c2}, \phi_3, \nu_{c3}]$, 
the objective function evaluates the negative of the total momentum transfer efficiency computed via the 
RCWA pipeline. By evaluating parallel populations and utilizing mutations and crossovers, the optimizer 
iteratively isolates the configuration that maximizes the efficiency.

The DE algorithm was configured with a population size of 40, 100 maximum iterations, 
and a spectral resolution of 500 frequency samples linearly spaced between 180 THz and 640 THz.
To ensure full spectral coverage across the target bandwidth without excessive inter-layer redundancy, 
the frequency search bounds are deliberately staggered. Each of these parameters are tuned within 
the given search space as summarized in Table \ref{tab:search space bounds}.

The simulation environment is governed by a set of global solver parameters listed in Table \ref{tab:rcwa_params}. 
The incident solar flux is evaluated across a selected spectral domain. 
The optical properties of the liquid crystal medium and the surrounding environmental index bounds 
are held constant throughout the optimization process. 
The guiding optimization performance metric is the magnitude of the transverse radiation pressure force
owing to the scattering of the blackbody solar spectrum. To evaluate this, the RCWA algorithm was 
executed until the optimized parameters (listed in Table \ref{tab:final_parameters}) were determined,
assuming normal incidence upon each layer.
Our numerical approach tracks the cascading transformations of 
the optical field from layer to layer.
That is, the spectral irradiance reaching Layer 2 is reduced by the reflectance of Layer 1, 
and the irradiance reaching Layer 3 is reduced by the combined reflectance of the preceding two layers. 

\begin{table}[htbp]
\centering
\caption{Search space bounds for the three-layer optimization. \smallskip}
\label{tab:search space bounds}
\begin{tabular}{cccc}
\hline
\noalign{\smallskip}
\shortstack{\textbf{Layer}} &
\shortstack{\textbf{Slant angle}\\$\phi$ \textbf{bounds}} &
\shortstack{\textbf{Cutoff Frequency} \\ \textbf{ ($\nu_c$) bounds}  }&
\shortstack{\textbf{Cutoff Wavelength} \\ \textbf{ ($\lambda_c$) bounds}  }\\
\hline
1 & $[15^{\circ}, 22^{\circ}]$ & $[350, 550]$ THz & [0.86, 0.54] $\mu$m \\
2 & $[15^{\circ}, 22^{\circ}]$ & $[200, 400]$ THz & [1.50, 0.75] $\mu$m\\
3 & $[15^{\circ}, 22^{\circ}]$ & $[100, 250]$ THz & [3.00, 1.20] $\mu$m\\
\hline
\end{tabular}
\end{table}

\begin{table}[htbp]
\centering
\caption{Global frequency and RCWA solver parameters utilized for the simulation. \smallskip}
\label{tab:rcwa_params}
\begin{tabular}{lc}
\hline
\noalign{\smallskip}
\textbf{Parameter} & \textbf{Value} \\
\hline
RCWA Frequency Sweep & $180 \text{ THz} - 640 \text{ THz}$ \\
Frequency Resolution & $500 \text{ steps}$ \\
Retained Harmonics ($M$) & $2$ (Total 5 modes: -2, -1, 0, 1, 2) \\
\hline
\end{tabular}
\end{table}

\section{Results}
A Differential Evolution (DE) optimization algorithm was employed to simultaneously tune the 
slant angle $\phi$ and cutoff frequency $\nu_c$ for each layer. 
The finalized primary design parameters are summarized in Table \ref{tab:final_parameters}.
Associated parameters for each layer include the transverse period 
$\Lambda_x = c/\nu_c = \lambda_c$, Bragg period 
$\Lambda_B = \Lambda_x \sin\phi$,
longitudinal period 
$\Lambda_z = \Lambda_B/\cos\phi$, and Bragg design wavelength
$\lambda_b = 2 \Lambda_B n_\text{eff} \cos \phi$.
These values are listed in Table \ref{tab:derived_geometry} along with the selected thickness values.
As expected the thickness of each layer is several times greater than the respective longitudinal period.
One of the advantages of Bragg gratings using high birefringent cholesteric liquid crystal polymer is the 
the ability to achieve high efficiency in thin films.  Here the total film thickness is $5.5 \; \mu$m, which
is remarkably thin for a three-layer diffractive solar sail.

\begin{table}[htbp]
\centering
\caption{Optimized primary design parameters for the three-layer Bragg stack.}
\label{tab:final_parameters}
\begin{tabular}{ccccc}
\hline
\noalign{\smallskip}
\shortstack{\textbf{Layer}} & \shortstack{\textbf{Thickness} \\ $L$} 
	& \shortstack{\textbf{Slant} \\ \textbf{Angle} \\ $\phi$}  
	& \shortstack{\textbf{Cutoff}  \\ \textbf{Frequency} \\ $\nu_c$} 
	& \shortstack{\textbf{Cutoff} \\  \textbf{Wavelength} \\ $\lambda_c = \Lambda_x$} \\
\hline
1 & $1.5 \mu\text{m}$ & $17.0^\circ$ & $435 \text{ THz}$ & 0.69 $\mu$m \\
2 & $2.0 \mu\text{m}$ & $17.5^\circ$ & $305 \text{ THz}$ & 0.98 $\mu$m\\
3 & $2.0 \mu\text{m}$ & $17.5^\circ$ & $206.5 \text{ THz}$ & 1.45 $\mu$m\\
\hline
\end{tabular}
\end{table}

\begin{table}[htbp]
\centering
\caption{Related parameters derived from optimized values of
$\phi$ and $\nu_c$.}
\label{tab:derived_geometry}
\begin{tabular}{cccccc}
\hline
\noalign{\smallskip}
\shortstack{\textbf{Layer}} & \textbf{$\Lambda_x$ ($\mu$m)} & \textbf{$\Lambda_z$ ($\mu$m)} & \textbf{$\Lambda_B$ ($\mu$m)} & \textbf{$\lambda_b$ ($\mu$m)}      & \textbf{$\nu_b$ (THz)}      \\
\hline
1  & 0.69 & 0.21 & 0.201 & 0.60 & 500 \\
2  & 0.98 & 0.31 & 0.296 & 0.87 & 344 \\
3  & 1.45 & 0.46 & 0.437 & 1.29 & 233 \\
\hline
\end{tabular}
\end{table}



\begin{figure}[htbp]
\centering\includegraphics[width=1\linewidth]{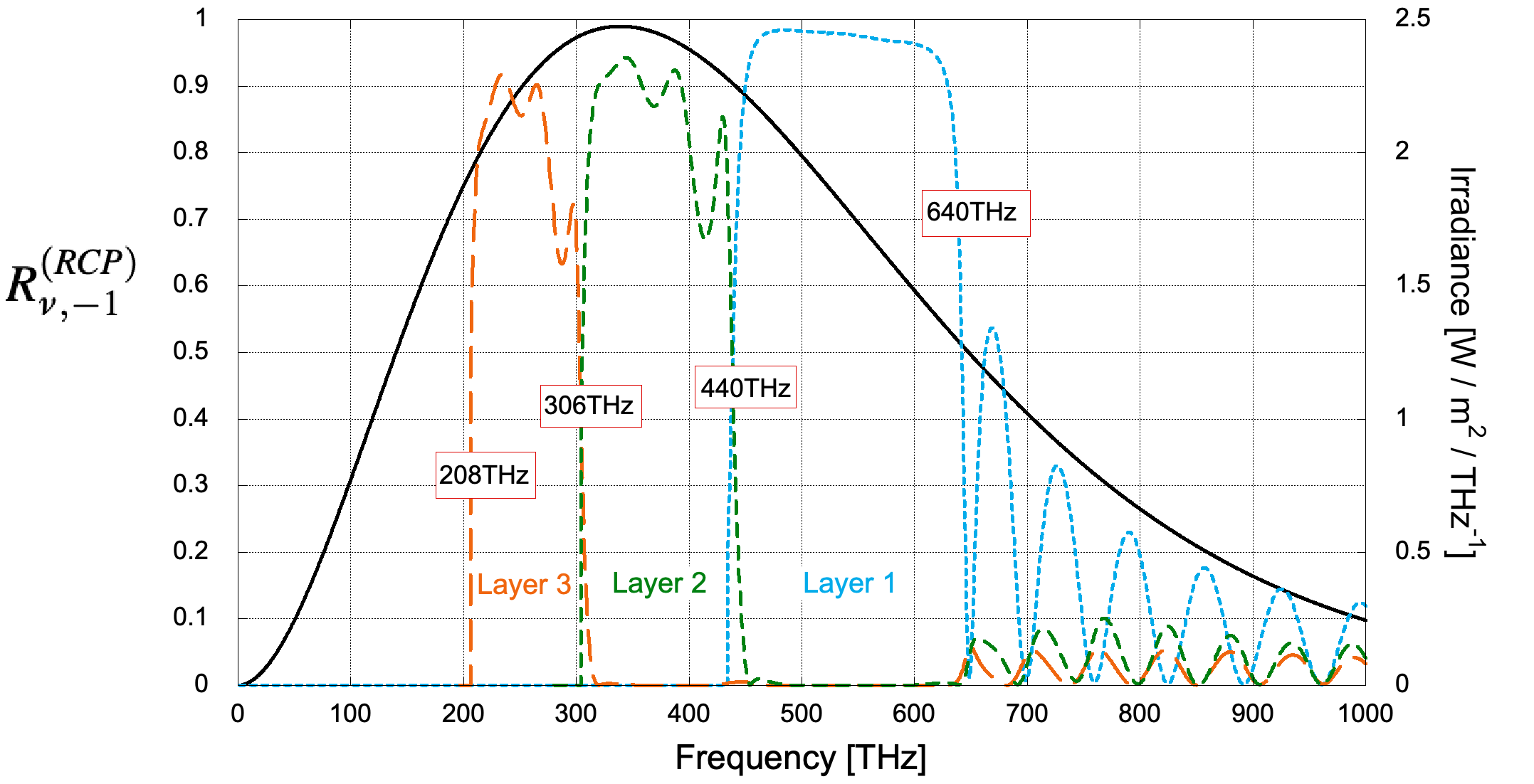}
\caption{Bragg reflection coefficients $R_{\nu,-1}$ for Layer 1 (blue line), Layer 2 (green line), and Layer 3 (orange line),
with the solar blackbody frequency spectrum (black line).}
\label{fig:combined_reflection}
\end{figure}

The reflectance spectra of the individual layers superimposed over the incident solar blackbody curve,
shown in Fig. \ref{fig:combined_reflection} for order $m=-1$, clearly shows high efficiency scattering
across three adjacent frequency bands, resulting in net broadband reflection nearly equal to the
full width half maximum blackbody spectrum.

The momentum transfer efficiency components are numerically calculated from the RCWA output,
assuming right circularly polarized (RCP) incident light, according to Eq.s \ref{eq:etax} and \ref{eq:etaz},
resulting in $\bar{\eta}_x^{(RCP)} = 25.6\%$ and $\bar{\eta}_z^{(RCP)} = 99.9\%$.
Although this value of transverse efficiency may be sufficient for station keeping at the Earth-Sun L1 point, as suggested
for a specular reflective sail at a $7^\circ$ tilt in Ref. \cite{Coco_2025},
a second layer would be needed to scatter both polarizations
away from the Earth for sun shading applications.  In the latter case the combined efficiency doubles to $\bar{\eta}_x^{(RCP)+(LCP)} = 51.2\%$.
Greater or lesser values of efficiency may be engineered by increasing or decreasing the net Bragg grating bandwidth.
What is more, the above design procedure may be readily adapted for a Bragg grating sail/sunshade that transmits
to Earth selective spectral bands that are important for photosynthesis (e.g., 600-680 nm or 440-500 THz) and photovoltaic power generation (e.g., 700-900 nm or 330 - 430 THz); for example, by eliminating Layer 2 in the above design
depicted in Fig. \ref{fig:combined_reflection}.

Further work, which is beyond the scope of this report,
will require specific sailcraft mass values and orbital optimization studies.  
Advantages of sun-facing diffractive sails over
canted specular reflection sails includes
reduced complexity for attitude and navigation control.
A simple roll-control maneuver (without changing the 
sun-facing attitude) provides navigational authority of a diffractive sail \cite{dubill2020attitude}. 

Finally we note that the radiation pressure force exerted upon
our sun-facing Bragg grating sail $A = 100$ m$^2$ would provide (for a single polarization state) 
a transverse force of 
$F_x = I_\text{sun} A \bar{\eta}_x^{(RCP)} / c = 0.012N$
and a longitudinal force of 
$F_z = I_\text{sun} A \bar{\eta}_z^{(RCP)} / c = 0.023N$.
These values would double for a two-stack film that
scatters both polarization states.

%
%

\section{Conclusion}
We theoretically designed and optimized a sun-facing Bragg grating solar sail comprised of a chiral liquid crystal polymer.  Using the Rigorous Coupled-Wave Analysis framework, we designed a three-layer Bragg reflector that scatters nearly 50\% of the solar blackbody spectrum, thereby providing a sufficient radiation pressure force for some space missions.  The designed grating operates on and primarily scatters into a single state of polarization. We suggest stacking two such Bragg gratings to achieve a useful radiation pressure force for missions such as sunshades at the Earth-Sun L1 point. As with all transportation systems, it is wise to engineer the vehicle for specfic tasks. Future work is therefore required to jointly optimize the Bragg grating sail with a specific sailcraft mission in mind.

\section{Acknowledgments}     
We thank the scientists and engineers at BEAM Engineering for Advanced Measurements Co. (Orlando, FL), particularly Aaron Becker for his assistance with our conversion of his anisotropic RCWA code from MATLAB to Python, and Nelson Tabiryan, Anna Tabiryan, and David Roberts for technical discussions related to fourth generation liquid crystal optics. We also thank Amber Dubill (Johns Hopkins Applied Physics Lab, Baltimore, MD) for technical discussions related to diffractive solar sails.  This work was supported by BEAM Co. via DARPA (award number 140D0423C0091/SUB-23682-01) and JHU/APL via NASA (award number 80MSFC22F0165/JHU APL 177864).







 \bibliographystyle{elsarticle-num}
\bibliography{references}

\end{document}